\documentclass[letterpaper]{aa}
\usepackage{graphicx}
\usepackage{txfonts}

\begin{document}

\title{On the local birth place of \object{Geminga}}

\author{L.~J. Pellizza\inst{1} \and R.~P. Mignani\inst{2} \and I.~A.
Grenier\inst{1,3} \and I.~F. Mirabel\inst{4}}

\institute{Service d'Astrophysique, DSM/DAPNIA, CEA Saclay, B\^at. 709, L'Orme
des Merisiers, 91191 Gif-sur-Yvette Cedex, France
\and European Southern Observatory, Karl Schwarzschild Strasse 2, D85748,
Garching b. Munchen, Germany
\and Universit\'e Paris VII Denis-Diderot, 2 place Jussieu, 75251 Paris Cedex
05, France
\and European Southern Observatory, Alonso de C\'ordova 3107, Vitacura, Casilla
19001, Santiago 19, Chile
}

\offprints{L.J. Pellizza, \email{leonardo.pellizza@cea.fr}}

\date{Received / Accepted}

\abstract{Using estimates of the distance and proper motion of \object{Geminga}
and the constraints on its radial velocity posed
by the shape of its bow shock, we investigate its
birth place by tracing its space motion backwards in time. Our results exclude
the \object{$\lambda$~Ori} association as the origin site because of the
large distance between both objects at any time. Our simulations place the
birth region at approximately 90--240~pc from the Sun, between $197\degr$ and
$199\degr$ in Galactic longitude and $-16\degr$ and $-8\degr$ in latitude, most
probably inside the \object{Cas-Tau} OB association or the 
\object{Ori~OB1}a association. We discard the possibility of the progenitor
being a massive field star. The association of \object{Geminga} with either
stellar association implies an upper limit of $M \approx 15\ M_{\sun}$ for the
mass of its progenitor. We also propose new members
for the \object{Cas-Tau} and \object{Ori~OB1} associations.

\keywords{stars: neutron -- pulsars: general -- pulsars: individual: Geminga
-- solar neighbourhood}}

\maketitle

\section{Introduction}
\label{intro}

The birth place of \object{Geminga} has been searched for since its optical
counterpart was identified and the first proper motion (Bignami et~al.
\cite{Big93}), spin-down age (Bignami \& Caraveo \cite{Big92}) and distance
estimates (Halpern \& Ruderman \cite{Hal93}) were obtained. Gehrels \& Chen
(\cite{Geh93}) proposed that the \object{Geminga} supernova event produced the
\object{Local Bubble}, while Frisch (\cite{Fri93}) argued that \object{Geminga}
was born somewhere in Orion. Smith et~al. (\cite{Smi94}) suggested the 
\object{$\lambda$~Ori} association (also known as \object{Collinder~69}) as the
most likely birth place, 450~pc away from the Sun. Moreover, the presence of an
\ion{H}{i} and dust ring surrounding \object{$\lambda$~Ori}, the size and
expansion velocity of which are consistent with the spin-down age of the pulsar
(Cunha \& Smith \cite{Cun96}), reinforced the association between
\object{Geminga} and this stellar group.

In recent years, the distance and proper motion of \object{Geminga}
($157^{+59}_{-34}$~pc and $170 \pm 4$~mas~yr$^{-1}$ respectively) were
accurately measured using HST images (Caraveo et~al. \cite{Car96}). As these
authors pointed out, \object{Geminga}'s radial velocity should be about 
$-700$~km~s$^{-1}$ in order to have reached its current position from the
\object{$\lambda$~Ori} association. This is a rather high value compared to
its transverse velocity of 126~km~s$^{-1}$. Using the {\it XMM-Newton}
Observatory, Caraveo et~al. (\cite{Car03}) succeeded in imaging the bow shock
produced by \object{Geminga} due to its motion through the ambient interstellar
medium. Modeling the bow shock shape yields an inclination of the 3D-velocity
to the plane of the sky that is smaller than $30\degr$. Given the transverse
velocity inferred from its proper motion and distance, this inclination
constrains the radial velocity to lower than 72~km~s$^{-1}$ in modulus,
therefore an order of magnitude lower than that needed to have reached the
current
position from \object{$\lambda$~Ori}. This fact prompted us to revisit the
potential birth place of \object{Geminga}.

\section{The \object{$\lambda$~Ori} association}
\label{lambdaori}

In order to analyze the possibility of \object{Collinder~69} being the birth
place of \object{Geminga}, we traced the space motion of both objects back in
time taking the spin-down age of the pulsar (0.342~Myr) as representative
of its true age. We neglected their acceleration in the Galactic potential,
since the pulsar age is much shorter than both of their orbital and epicyclic
periods. We also neglected possible changes in velocity produced by close
encounters with other stars, because the stellar density in the solar
neighbourhood and the velocity of \object{Geminga} imply a mean time between
encounters more than ten orders of magnitude greater than the pulsar age. We
computed the position of both \object{Geminga} and \object{Collinder~69} for
the last 0.342~Myr at intervals of 0.01~Myr, together with a full covariance
matrix for them. The distance $d$ between these objects at each timestep and
its uncertainty $\epsilon$ were also computed. The uncertainty was derived from
the covariance matrices of the positions, and takes into account the errors on
all measured parameters (\object{Geminga} distance and proper motion and 
\object{Collinder~69} distance, proper motion and radial velocity).

\begin{figure}
\resizebox{\hsize}{!}{\includegraphics{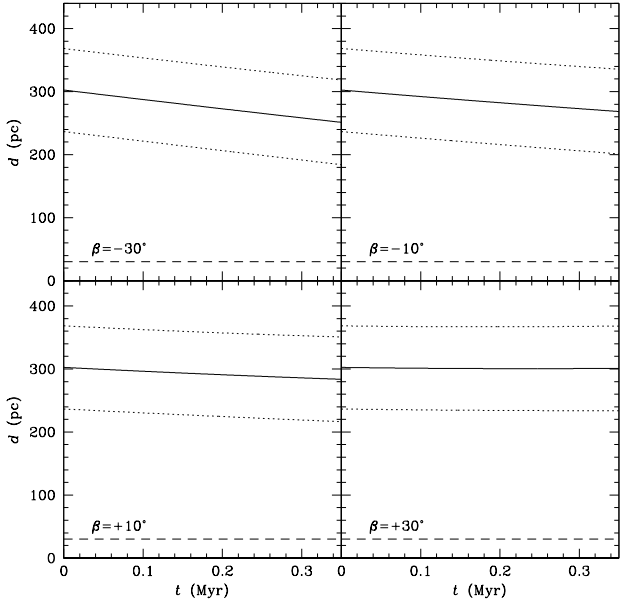}}
\caption{Distance between \object{Geminga} and \object{Collinder~69} as a
function of look-back time for four values of the inclination $\beta$ of the
pulsar velocity onto the plane of the sky. Solid lines give the mean
distances, while dotted lines indicate the error limits. These errors account
for all the uncertainties in both \object{Geminga} and \object{Collinder~69}
measured parameters. Dashed lines indicate the radius of \object{Collinder~69}.
This figure shows that the distance between these objects remained much greater
than the radius of the cluster for the last 0.342~Myr.}
\label{Col_69_dist}
\end{figure}

For \object{Collinder~69} we took the coordinates
($\alpha = 5^{\mathrm h} 35^{\mathrm m} 06\fs 0$, $\delta = +9\degr 56\arcmin
00\arcsec$), proper motion ($\mu_\alpha \mathrm{cos}\,\delta = 0.45 \pm
2.80$~mas~yr$^{-1}$,
$\mu_\delta = -2.40 \pm 2.80$~mas~yr$^{-1}$) and radial velocity ($30.5 \pm
2.6$~km~s$^{-1}$) from the open cluster catalogue of Dias et~al.
(\cite{Dia02}). Its current distance ($450 \pm 50$~pc) was taken from Dolan \&
Mathieu (\cite{Dol01}). For \object{Geminga}, we used the coordinates 
($\alpha = 6^{\mathrm h} 33^{\mathrm m} 54\fs 15$, $\delta = +17\degr 46\arcmin
12\farcs 9$) from Caraveo et~al. (\cite{Car98}), the parallax ($6.36 \pm
1.74$~mas) and proper motion ($\mu_\alpha \mathrm{cos}\,\delta = 138 \pm
4$~mas~yr$^{-1}$, $\mu_\delta = 97 \pm 4$~mas~yr$^{-1}$) measured by Caraveo
et~al.
(\cite{Car96}), and a set of four values of the angle $\beta$ between its
velocity and the plane of the sky ($\beta = -30\degr$, $-10\degr$, $+10\degr$
and $+30\degr$) which satisfy the constraints on the bow shock.

Figure~\ref{Col_69_dist} presents the distance $d$ between \object{Geminga} and
\object{Collinder~69} as a function of look-back time $t$ for four values of
$\beta$. This figure clearly shows that during the whole time interval, the
radius of \object{Collinder~69} ($R=30$~pc, Dolan \& Mathieu \cite{Dol01}) is
much smaller than the distance between \object{Geminga} and this cluster.
Hence, it is very unlikely that this stellar association is the birth place of
\object{Geminga}.

\begin{figure}
\resizebox{\hsize}{!}{\includegraphics{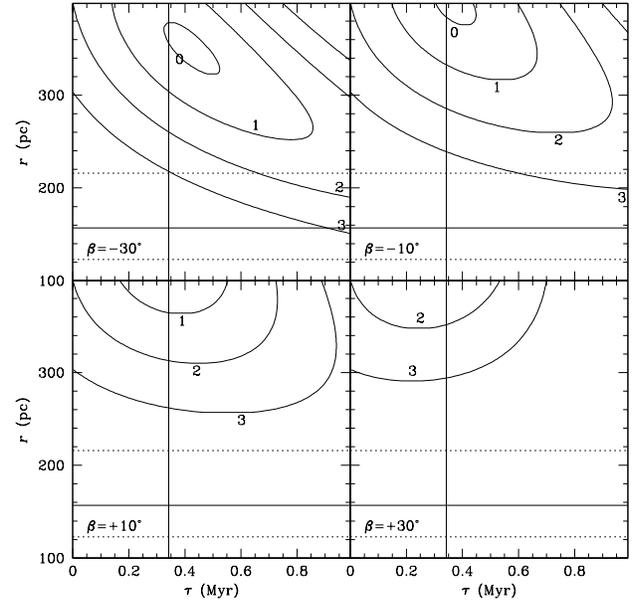}}
\caption{Contours of the ratio $f$ of the separation $d_\mathrm{birth}$ of
\object{Geminga} and the \object{Collinder~69} outer boundary at pulsar
birth to its uncertainty $\epsilon_\mathrm{birth}$. The contours are labeled
by the value of $f$. The vertical and horizontal solid lines respectively
indicate the spin-down age and the mean current distance of the pulsar given
by parallax measurements. Dotted lines indicate the current distance error. The
region inside the $f = 1$ contour, for which the association is possible, is at
variance with the current data.}
\label{Col_69_age_dis}
\end{figure}

To investigate the effects of possible wrong distance/age estimates, we 
evolved \object{Geminga} backwards in time from different current distances
$r$ and computed the separation at birth $d_\mathrm{birth}$ between the
pulsar and the cluster outer boundary nearest to it, and its uncertainty
$\epsilon_\mathrm{birth}$, both as a function of $r$ and age $\tau$. In this
case, $r$ is considered a parameter with no error, hence
$\epsilon_\mathrm{birth}$ does not take into account the current
distance error. Contours of the function $f = d_\mathrm{birth} /
\epsilon_\mathrm{birth}$ are displayed in Fig.~\ref{Col_69_age_dis} for four
values of $\beta$. It shows that a pulsar born within the cluster ($f \leq 1$)
would be much more distant than allowed by the parallax measurements and bow
shock shape for any age. Hence, the relationship between \object{Geminga} and
the \object{$\lambda$~Ori} association is very unlikely, unless the cluster
distance and/or the pulsar distance/age are seriously revised. Specifically, a
pulsar current distance of almost twice the parallax distance and/or an age
of twice the spin-down age would be needed to make the association possible for
$\beta = -30\degr$. However, these values are unlikely because other pieces of
evidence, such as the agreement between X-ray or optical/UV data and
theoretical models (Halpern \& Ruderman \cite{Hal93}; Bignami et~al.
\cite{Big96}), also favour the spin-down age and parallax distance.

\section{Nearby associations}
\label{nearbyassoc}

We searched for other suitable birth places for \object{Geminga} near the
position it had 0.342~Myr ago. This position was very close, less than 200~pc
from the Sun for $|\beta| \leq 30\degr$, so a potential stellar cluster would
have a large angular scale making it difficult to determine its structure. The
method applied formerly to \object{Collinder~69} is not well suited for these
associations, as their centers and boundaries are rather uncertain. Hence, we
studied the stellar associations in the solar neighbourhood whose stars are
grouped in position and velocity (De Zeeuw et~al. \cite{DeZ99}) and we directly
compared the stellar positions with the likely birth place of \object{Geminga},
defined as the error box of its position 0.342~Myr ago. Given $|\beta| \leq
30\degr$ and the current distance and proper motion uncertainties, the likely
birth place is contained in a box defined by $197\degr < l < 199\degr$,
$-16\degr < b < -8\degr$ and $90\ \mathrm{pc} < r < 240\ \mathrm{pc}$, where
$(l,b)$ are the usual Galactic coordinates and $r$ is the distance from the
Sun.

We selected for our analysis all the O and B stars in a slightly greater box,
defined by $190\degr < l < 205\degr$, $-45\degr < b < 0\degr$ and $r < 400$~pc.
The coordinates, parallaxes and proper motions of the stars in this box were
taken from the All-Sky Compiled Catalogue of 2.5 million stars (ASCC2.5) of
Kharchenko (\cite{Kha01}). This catalogue has a limiting $V$ magnitude of
12--14, much greater than the apparent magnitude of the faintest B stars if
located at 400~pc from the Sun ($V \approx 8$). Hence, we expect no photometric
selection effects in our sample. As we
need stars with accurate positions to trace the stellar associations, we
included in our sample only three groups of stars: 1) stars with accurate
Hipparcos parallaxes ($\delta \pi / \pi < 0.25$); 2) stars with accurate
distance moduli derived from Walraven photometry by Brown et~al.
(\cite{Bro94}); and 3) stars for which good photometric distances can be
computed from the data in the ASCC2.5 catalogue (i.e. from MK spectral types,
luminosity classes and Johnson $V$ magnitudes and $B-V$ colours). To compute
the stellar distances in the third group, we used the MK calibration of
Schmidt-Kaler (\cite{Sch82}). Stars that do not comply with any of these
conditions were rejected. The selected region contains 181 useful O and B
stars, that consist of 83 stars with accurate trigonometric parallaxes, 122
with distance moduli from Brown et~al. (\cite{Bro94}) and 38 with photometric
distances computed from ASCC2.5 data.

The stellar distances computed from ASCC2.5 data take into account the
interstellar extinction derived from the comparison between observed and
intrinsic stellar colours. A correct determination of the extinction being
crucial for a good distance estimate, we compared the observed stellar
reddening $E(B-V)_\mathrm{o}$ with that derived at the same position in the sky
from a Galactic dust reddening map, $E(B-V)_\mathrm{d}$ (Schlegel et~al.
\cite{Sch98}). Away from the Galactic plane, as is the case near Orion,
both values should reasonably agree and the dust reddening should give an
upper limit to the observed one. Hence, when $E(B-V)_\mathrm{o} >
E(B-V)_\mathrm{d}$ we replaced the former with the latter. In the opposite case
we used $E(B-V)_\mathrm{o}$ even for stars that show a large difference
between both reddening values, as these are probably very nearby stars. Note
that replacing the observed colour excess with the dust-derived one in these
stars would make their computed distances even smaller, thus making more
improbable the higher value of extinction obtained from dust maps.

Using either parallax or photometric distances, we computed the Galactic
$(X,Y,Z)$ Cartesian coordinates of the sample stars. Only two of the nearby OB
associations listed by De Zeeuw et~al. (\cite{DeZ99}) were found to have stars
within the box of interest for \object{Geminga}; these are the
\object{Cas-Tau} and the \object{Ori~OB1} associations. A subset of
our sample of stars has not only good distances but also good radial velocity
determinations. In these cases, taking the radial velocities from the Catalogue
of Radial Velocities with Astrometric Data (Kharchenko et~al. \cite{Kha04}), we
computed the stellar velocities $(U,V,W)$. The velocities were corrected for
the solar motion using the values $U_{\sun} = 9.7$~km~s$^{-1}$, $V_{\sun} =
5.2$~km~s$^{-1}$ and $W_{\sun} = 6.7$~km~s$^{-1}$ (Bienaym\'e \cite{Bie98}),
and for Galactic rotation using the Oort constants $A =
14.8$~km~s$^{-1}$~kpc$^{-1}$ and $B = -12.4$~km~s$^{-1}$~kpc$^{-1}$. These
velocities were used to identify previously unknown members of the OB
associations.

\subsection{The \object{Cas-Tau} association}
\label{castau}

The \object{Cas-Tau} association is a very nearby association
covering an area of approximately $100\degr \times 60\degr$ in the sky, and
extending between 125~pc and 300~pc from the Sun. Only 9 of the 83 O and B
stars with membership probabilities greater than 50\% listed by De Zeeuw et~al.
(\cite{DeZ99}) are found in our box. Since the ASCC2.5 contains more stars
than the Hipparcos catalogue used by De Zeeuw et~al. (\cite{DeZ99}), we
searched for possible new \object{Cas-Tau} members inside our box. For this
purpose, we compared the velocity distributions in our sample with those of all
\object{Cas-Tau} members listed by De Zeeuw et~al. (\cite{DeZ99}) with
available radial velocities. The mean heliocentric \object{Cas-Tau} velocity
components are $U_\mathrm{CT} = -13.24$~km~s$^{-1}$, $V_\mathrm{CT} =
-19.69$~km~s$^{-1}$ and $W_\mathrm{CT} = -6.38$~km~s$^{-1}$ (De Zeeuw et~al.
\cite{DeZ99}). Using the solar motion relative to the LSR given by Bienaym\'e
(\cite{Bie98}), the \object{Cas-Tau} velocity relative to the LSR is
$U_\mathrm{CT,LSR} = -3.54$~km~s$^{-1}$, $V_\mathrm{CT,LSR} =
-14.49$~km~s$^{-1}$ and $W_\mathrm{CT,LSR} = +0.32$~km~s$^{-1}$.

\begin{figure}
\resizebox{\hsize}{!}{\includegraphics{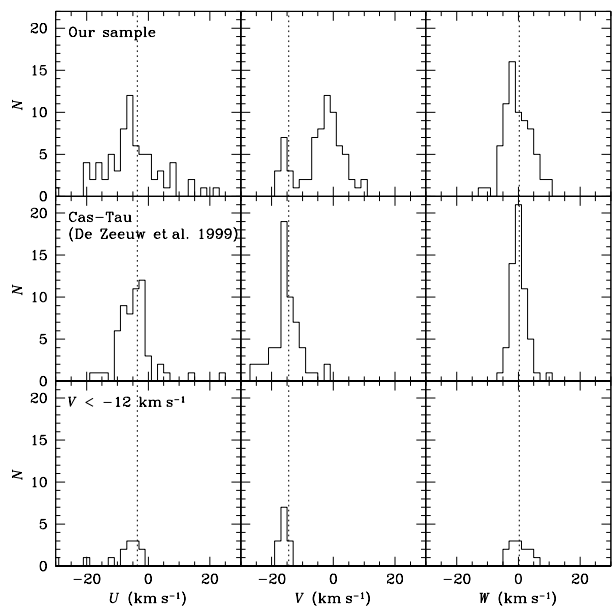}}
\caption{$(U,V,W)$ velocity distributions for all stars in our sample (upper
row), for the \object{Cas-Tau} stars identified by De Zeeuw et~al.
(\cite{DeZ99}) in and out of our box (middle row), and for the stars from
the most negative $V$ peak in our sample ($V < -12$~km~s$^{-1}$, lower row).
These last 13 stars include 4 known \object{Cas-Tau} members, 8 stars which we
propose as new \object{Cas-Tau} members on the basis of their position and
velocity, and 1 background field object. Dotted lines show the mean $(U,V,W)$
of the \object{Cas-Tau} association. All velocities are relative to the LSR.}
\label{CasTauVel}
\end{figure}

\begin{figure*}
\resizebox{\hsize}{!}{\includegraphics{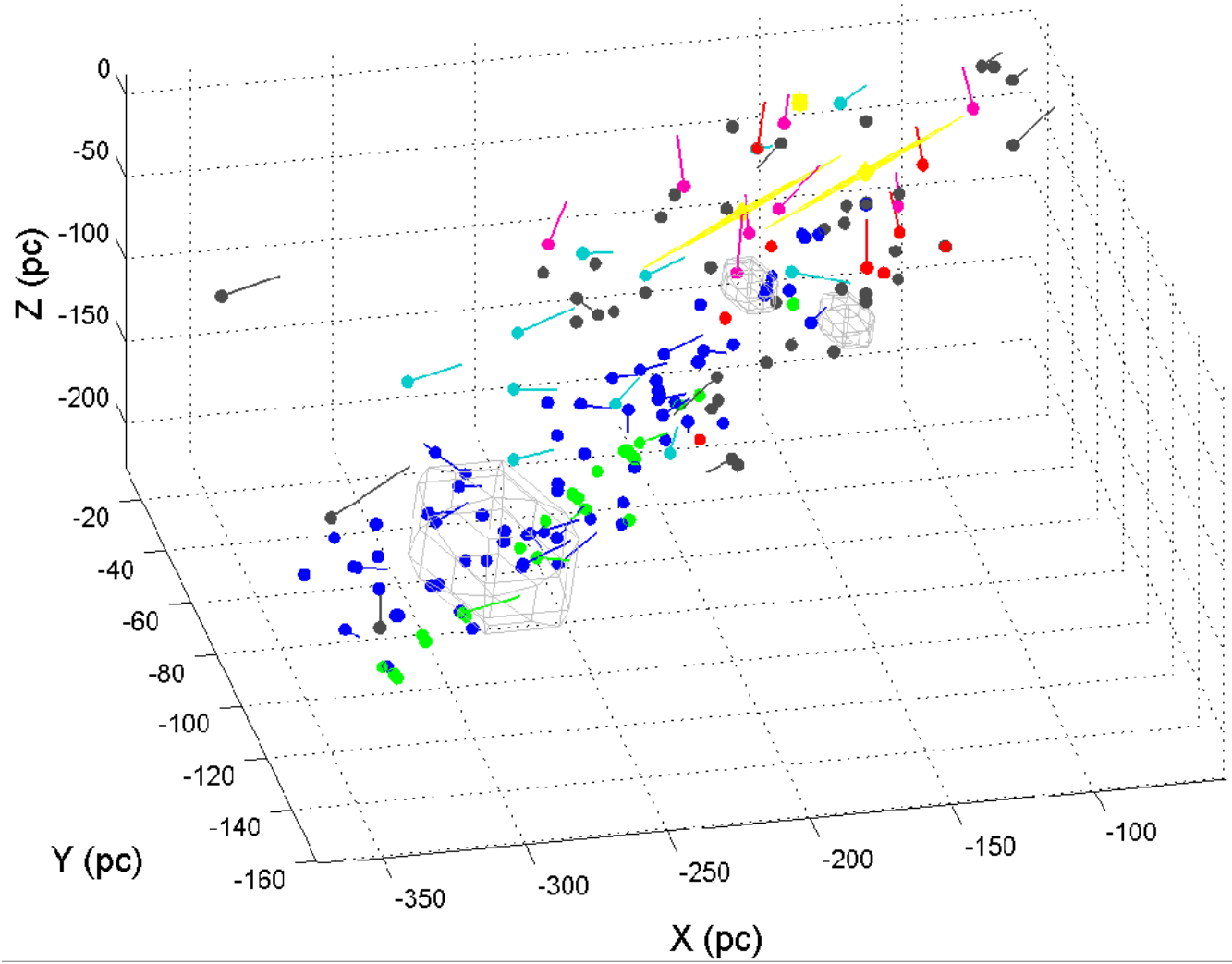}}
\caption{Spatial distribution of the stars (filled circles) belonging to the
\object{Cas-Tau} (red), \object{Ori~OB1}a (dark-blue) and \object{Ori~OB1}b/c
(green) associations. Pink and light-blue circles outline the proposed new
members of the \object{Cas-Tau} and \object{Ori~OB1}a associations,
respectively, while gray ones represent field stars and stars for which the
lack of data prevents any classification. The current position of
\object{Geminga} is marked by the yellow star, while yellow diamonds surrounded
by ellipses indicate the positions at birth and likely birthplaces for
$\beta = -30\degr$ (left one) and $\beta = +30\degr$ (right one). The sticks
give the stellar velocities when available, their lengths corresponding to the
distance traveled in the last 1~Myr. The skeletal spheres, that appear
elongated due to projection effects, mark the main location of
\object{Ori~OB1}a (the large one near the lower left corner) and the two more
nearby subgroups identified by Platais et al. (\cite{Pla98}) as part of
\object{Ori~OB1}a (the small ones near the upper right corner). For all
possible values of $|\beta| \leq 30\degr$, the birth place of \object{Geminga}
falls among \object{Cas-Tau} stars and near the edge of \object{Ori~OB1}a, or
inside it if it is extended to the proposed new members.}
\label{field}
\end{figure*}

As Fig.~\ref{CasTauVel} shows, the $U$ and $W$ distributions of
\object{Cas-Tau} stars are quite similar to those in our sample. A difference
arises in the $V$ distribution, which is single-peaked for \object{Cas-Tau}
stars but presents two peaks in our sample. The peak around $-2$~km~s$^{-1}$
is not present for \object{Cas-Tau} stars, hence it contains stars not
belonging to this association. The most negative peak corresponds to the
\object{Cas-Tau} distribution and the 13 stars in this peak can be selected by
the condition $V < -12$~km~s$^{-1}$. Four of them are indeed known
\object{Cas-Tau} members from De Zeeuw et~al. (\cite{DeZ99}), while one is too
far away to belong to this group. The velocity distribution of the others, as
well as their 3D-position displayed in Fig.~\ref{field}, strongly suggest
their belonging to the association. So, we propose these 8 stars as new
\object{Cas-Tau} members\footnote{Their ASCC2.5 catalogue numbers are 845\,656,
925\,805, 929\,279, 1\,013\,063, 1\,013\,660, 1\,016\,668, 1\,017\,786 and
1\,192\,880.}.

Figure~\ref{field} shows the position of the stars in our box according to
their membership, together with the confidence regions for the position of
\object{Geminga} at birth (i.e. 0.342~Myr ago) obtained from the simulations.
This figure clearly shows that, for all values of $\beta$ consistent with the
bow-shock data, the birth place of \object{Geminga} is located well inside the
\object{Cas-Tau} association. Hence, \object{Cas-Tau} is a very
likely candidate for the parent association.

\subsection{The \object{Ori~OB1} association}
\label{oriob1}

The other association contained in our box is the \object{Ori~OB1} association,
composed of four subgroups named \object{Ori~OB1}a to \object{Ori~OB1}d (De
Zeeuw et al. \cite{DeZ99}). Our box contains 86 of the 132 O and B members of
\object{Ori~OB1}a, 25 of the 84 members of \object{Ori~OB1}b, only 1 out of 126
in \object{Ori~OB1}c, and none of the \object{Ori~OB1}d massive stars (see
Fig.~\ref{field}). The \object{Ori~OB1}b/c stars in our box are located near
the high-longitude edge, far away from the possible birth places. Hence, the
only subgroup possibly associated with \object{Geminga} is \object{Ori~OB1}a.

The majority of \object{Ori~OB1}a stars in Fig.~\ref{field} cluster in two
subgroups, located at approximately 240~pc and 340~pc from the Sun. The first
one is close to the position of two new stellar groups found by Platais et~al.
(\cite{Pla98}) in the Hipparcos data, which they propose to be part of
\object{Ori~OB1}a. The second one corresponds to the main position of
\object{Ori~OB1}a given by De Zeeuw et~al. (\cite{DeZ99}), and is also found
in the list of Platais et~al. (\cite{Pla98}). Hence, \object{Ori~OB1}a appears
as a highly elongated system, with a $5\degr$ radius in the plane of the sky
and extending from 200~pc to more than 400~pc from the Sun. Although such
apparent radial extension could stem from distance uncertainties, in particular
in the correction of interstellar extinction, this elongated geometry seems to
be real. Using the different distance estimates for the stars which have more
than one, we always reach the same conclusion, namely that
\object{Ori~OB1}a extends at least between 200~pc and 400~pc from the Sun.

\begin{figure}
\resizebox{\hsize}{!}{\includegraphics{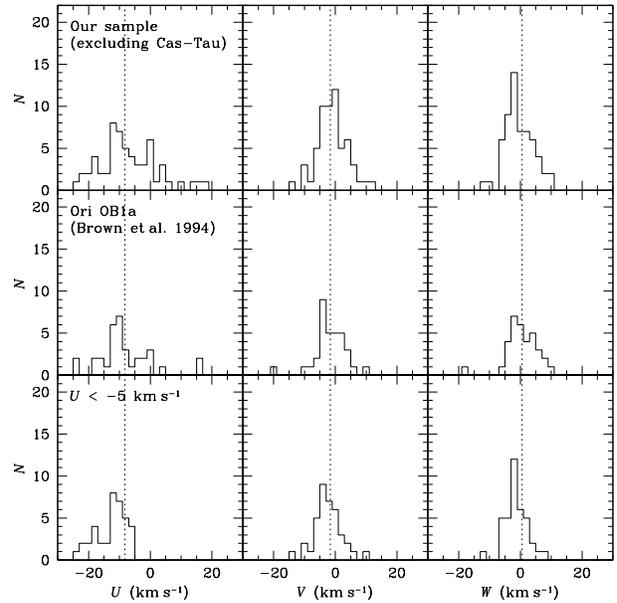}}
\caption{$(U,V,W)$ velocity distributions for all stars in our sample but in
the \object{Cas-Tau} association (upper row), for the known \object{Ori~OB1}a
stars in and out of our sample (middle row), and for the stars with $U <
-5$~km~s$^{-1}$ in our sample (lower row). These 38 stars include 19 known
\object{Ori~OB1}a members, 3 \object{Ori~OB1}b members and 16 stars sharing the
same motion as the \object{Ori~OB1}a ones, 11 of which are located along the
elongated stream, the others being either far from it or much closer to the
Sun. Dotted lines show the mean $(U,V,W)$ of the \object{Ori~OB1}a association.
All velocities are relative to the local LSR of each star.}
\label{OriVel}
\end{figure}

Figure~\ref{field} displays the velocity vectors for those stars having radial
velocity measurements. Their lengths correspond to the distance traveled in the
last 1~Myr. Figure~\ref{OriVel} shows that the whole \object{Ori~OB1}a
association
moves almost radially away from the Sun, with nearly null $V$ and $W$
components, but an appreciable negative $U$ component with a peak near
$U \approx -10$~km~s$^{-1}$. On the other hand, field stars rotating with the
Galaxy should not deviate appreciably from a circular orbit and should have a
close to null velocity relative to their own LSR. The $U$ distribution of the
stars in our box extends to large negative values because of the presence of
\object{Ori~OB1}a members. The smaller sample with $U < -5$~km~s$^{-1}$
displayed in the lowest row of Fig.~\ref{OriVel} exhibits velocities in good
agreement with those of \object{Ori~OB1}a. They include 19 known
\object{Ori~OB1}a members and 11 possible new ones\footnote{The ASCC2.5
catalogue numbers of the proposed new members are 846\,100, 924\,356, 924\,551,
1\,010\,325, 1\,012\,501, 1\,013\,386, 1\,013\,408, 1\,103\,267, 1\,104\,520,
1\,105\,105 and 1\,194\,478.}. In Fig.~\ref{field} they appear to be
scattered within both subgroups, near the lower longitude edge of the dataset.
Five of them indicate a possible extension of the association on the near side,
to 140~pc from the Sun. Given the long radial extension of the association and
the almost radial velocities of its members, \object{Ori~OB1}a appears as an
unusual vast stream of stars. How it could have been triggered by the expanding
wave of the \object{Gould Belt} is unclear (Perrot \& Grenier \cite{Per03}).

Regarding the origin of \object{Geminga}, Fig.~\ref{field} shows that the
near end of the \object{Ori~OB1}a association (without considering the possible
new members) is close to but does not overlap the birth place of the pulsar,
hence making the probability of the progenitor of \object{Geminga} belonging to
\object{Ori~OB1}a rather low. However, if the possible new members are
included, the association extends well over the birth region of
\object{Geminga}. In this case, \object{Geminga} would have been born in a
region where the two large \object{Cas-Tau} and \object{Ori~OB1}a associations
intersect, making it impossible to decide, by kinematical means, which is the
parent one.

\subsection{Nearby field stars}
\label{fieldstars}

It is also possible that the \object{Geminga} progenitor is a field star. Nine
stars in our sample do not belong to any association and another 27 have
neither a previous classification nor radial velocity data, thus making it
impossible to know if they belong to an association or not. The earliest
main-sequence stars in both groups have spectral type B8V. Their mean mass of
approximately 4~$M_{\sun}$ is far lower than the minimum mass threshold for
supernova explosion (7--8~$M_{\sun}$), and they have lifetimes longer than
150~Myr on the main sequence. Hence, if our small sample is representative of
the field population around the birth place of \object{Geminga}, this
population is much older than that of the local OB associations, and has no
stars massive and young enough to account for the production of a pulsar in the
last 0.342~Myr.

\subsection{The mass of the progenitor of \object{Geminga}}
\label{massprog}

If \object{Cas-Tau} is the parent association of \object{Geminga},
a constraint can be derived on the mass of its progenitor. This association has
an age $\tau_\mathrm{CT} \approx 50$~Myr (De Zeeuw et~al. \cite{DeZ99}). At
this age, stars evolving out of the main sequence now (and also at the birth
time of \object{Geminga}, since $\tau_\mathrm{CT} \gg 0.342$~Myr) have a mass
of the order of 7~$M_{\sun}$, marginally consistent with the current
theoretical limits for type-II supernovae (hereafter SNII) progenitors of
7--8~$M_{\sun}$. A study of the individual \object{Cas-Tau} stars present in
our box shows, on the other hand, that the earliest main-sequence stars are a
B3V and possibly a B2IV-V star. Following the calibration of Drilling \&
Landolt (\cite{Dri99}), these stars should have 8--10~$M_{\sun}$, a mass range
consistent with the theoretical limits for SNII progenitors. A similar mass
range is obtained from the mean masses of B2V and B3V stars in the catalogue of
Belikov (\cite{Bel95}). According to these data, the \object{Cas-Tau} age would
be in the range 25--37~Myr, somewhat younger than the estimate of De Zeeuw
et~al. (\cite{DeZ99}). Hence, if born in the \object{Cas-Tau}
association, the \object{Geminga} progenitor would have been among the least
massive stars capable of producing a SNII (7--10~$M_{\sun}$).

On the other hand, the \object{Ori~OB1}a association has a young age
$\tau_\mathrm{Ori~OB1a} = 11.4 \pm 1.9$~Myr (Brown et~al. \cite{Bro94}), which
implies a mass of approximately 15~$M_{\sun}$ for the stars leaving the main
sequence now (and 0.342~Myr ago). This conclusion is supported by the fact that
the earliest \object{Ori~OB1}a main-sequence stars in our sample are of
spectral type B1V, which have mean masses of 12--15~$M_{\sun}$ (Drilling \&
Landolt \cite{Dri99}, Belikov \cite{Bel95}), large enough to produce type II
supernovae.

Given the uncertainty in the parent association between \object{Cas-Tau} and
\object{Ori~OB1}a and the fact that there are no main sequence stars with $M >
15\ M_{\sun}$ in the potential birth region of the pulsar, we can derive a
robust upper limit of  15~$M_{\sun}$ for the progenitor mass.

\section{Conclusions}
\label{conc}

Using the most recent values and constraints for positions, distances, proper
motions and radial velocities of \object{Geminga} and the
\object{$\lambda$~Ori} association (\object{Collinder~69}), we simulated
their space motions for the last 0.342~Myr (the spin-down age of the pulsar)
and discarded the hypothesis that \object{Geminga} was born in this association.

Our simulations locate the most likely birth place of the pulsar between
approximately $d = 90 - 240$~pc from the Sun, $197\degr$ to $199\degr$ in
Galactic longitude, and $-16\degr$ to $-8\degr$ in Galactic latitude. A search
for the possible
parent groups of \object{Geminga} within this region led us to local OB
associations. Among these, the \object{Cas-Tau} one, a
$\approx 50$~Myr-old group, emerges as the most likely parent association. The
birth place of \object{Geminga} also lies very near the edge of the
\object{Ori~OB1}a association, a much younger group ($11.4 \pm 1.9$~Myr). If
the proposed elongated geometry for this association is confirmed, namely that
it constitutes a stream of stars moving almost radially away at distances
between 140 and 400~pc from the Sun, then the birth place of \object{Geminga}
would also be inside \object{Ori~OB1}a. It would be in the
intersection region between \object{Ori~OB1}a and \object{Cas-Tau}.
The possibility of the progenitor of \object{Geminga} being a field star is
unlikely because of their older age in this region. Further constraints on the
\object{Geminga} birth place could come from X-ray observations with {\it
Chandra}, which could better constrain
the inclination of the 3D-velocity onto the plane of the sky. However, because
of the large dimensions of \object{Cas-Tau} and \object{Ori~OB1}a, the better
localization of the birth place would not alter our conclusions about the
parent association.

The origin of the \object{Local Bubble} in the \object{Geminga} SN explosion
(Gehrels \& Chen \cite{Geh93}) is not confirmed by our results. The closest
birth place for \object{Geminga} is found near the Bubble edge and the other
possible sites extend much beyond the Bubble wall (Lallement et~al.
\cite{Lall03}). Recent investigations also show that the \object{Local Bubble}
could not have been formed by a single SN event, but by multiple SNe which
would have taken place in the \object{Pleiades}~B1 moving group (Bergh\"ofer \&
Breitschwerdt \cite{Ber02}).

We find similar constraints on the \object{Geminga} progenitor mass in both
parent associations. It is no greater than 7--10~$M_{\sun}$ in
\object{Cas-Tau}, placing it among the least massive stars capable of producing
a pulsar. The mass limit is slightly higher, 12--15~$M_{\sun}$, in the younger
\object{Ori~OB1}a. $M = 15 M_{\sun}$ is a robust upper limit for the
progenitor mass in this region.

\begin{acknowledgements}

We would like to thank Dr. Marc Rib\'o for a careful reading of the manuscript
and useful comments on it.

\end{acknowledgements}

\end{document}